# NFV Based Gateways for Virtualized Wireless Sensor Networks: A Case Study

Carla Mouradian[¥1], Tonmoy Saha[¥2], Jagruti Sahoo[¥3], Roch Glitho[¥4], Monique Morrow[€1] , Paul Polakos[£1]

[¥]Concordia University, Montreal, Canada,

[€]CISCO Systems, Zurich, Switzerland,

[£]CISCO Systems, New Jersey, USA

[¥1]ca_moura@encs.concordia.ca, [¥2]to_saha@encs.concordia.ca, [¥3]jagrutiss@gmail.com,
[¥4]glitho@ece.concordia.ca, [€1]mmorrow@cisco.com, [£1]ppolakos@cisco.com

*Abstract*— **Virtualization enables the sharing of a same wireless sensor network (WSN) by multiple applications. However, in heterogeneous environments, virtualized wireless sensor networks (VWSN) raise new challenges such as the need for on-the-fly, dynamic, elastic, and scalable provisioning of gateways. Network Functions Virtualization (NFV) is an emerging paradigm that can certainly aid in tackling these new challenges. It leverages standard virtualization technology to consolidate special-purpose network elements on top of commodity hardware. This article presents a case study on NFV based gateways for VWSNs. In the study, a VWSN gateway provider operates and manages an NFV based infrastructure. We use two different brands of wireless sensors. The NFV infrastructure makes possible the dynamic, elastic, and scalable deployment of gateway modules in this heterogeneous VWSN environment. The prototype built with Openstack as platform is described.**

## I. INTRODUCTION

In recent years, virtualization of wireless sensor networks [3]-[5] has drawn significant attention from both academia and industry. Virtualization technology abstracts the sensor resources as logical units and allows their efficient usage by multiple concurrent applications, even with conflicting requirements and goals. New applications can be deployed in the same WSN with minimal efforts. More importantly, re-use of the same sensors by multiple applications transforms WSN into a multi-purpose sensing platform, where several virtual WSNs (VWSNs) are created on-demand, each tailored for a specific task or objective.

A novel architecture for VWSN that uses constrained application protocol (CoAP) is proposed in [5], wherein the potential benefits of virtualization are illustrated by using a scenario where a single WSN is shared by multiple applications; one of which is a fire monitoring application. In order to send sensor measurements and receive monitoring commands to and from applications, the dynamically formed VWSNs must have access to WSN gateway, especially in heterogeneous wireless sensor environments. In a multi-vendor WSN, the proprietary network technologies are often incompatible with each other, leading to increased complexity of WSN gateway. Furthermore, it is difficult and much expensive to upgrade the WSN gateway with deployment of new brand of sensors. The network processing capability of traditional WSN gateways cannot scale when the number of applications and the corresponding workload in VWSN changes dynamically.

Network Functions Virtualization (NFV) [1], [2] is an emerging paradigm that can certainly aid in addressing the aforementioned challenges. NFV leverages standard virtualization technology to consolidate dedicated network elements such as home gateways, firewalls, network address translation (NAT), and load balancers, on top of commodity hardware. By implementing network functions in software modules termed as virtual network functions (VNFs), NFV aims at reducing the operational cost and providing hardware-independence. Its key benefits include on-the-fly, dynamic, scalable, and elastic provisioning. The key benefits of virtualization including fast provisioning, Virtual Machine (VM) mobility, etc. are also achieved in NFV environments.

This paper presents a case study on NFV based gateways for VWSN. In the study, we introduce a new business entity, the VWSN gateway provider, in addition to the traditional roles of application provider and VWSN provider. The case study involves two VWSN providers that use sensors of different brands (e.g. Java Sun Spot and Advanticsys). In this study, we design VNFs for key modules of WSN gateway such as protocol conversion and information model processing. These VNFs are instantiated on the fly and chained to achieve elastic and scalable provisioning of gateway modules. The VWSN gateway provider operates and manages an NFV based infrastructure that enables the deployment of VNFs in VWSN domains.

The remainder of this paper is organized as follows. The next section introduces motivating scenario, requirements, and discusses state-of-the-Art. The proposed architecture is presented in the Section 3. Section 4 presents the implementation details along with the prototype. In the last section, we conclude the paper and outline our future works.

## II. CRITICAL OVERVIEW OF STATE-OF-THE-ART

### A. Motivating Scenario

The ability of sensors to sustain in harsh environments makes WSN a potential tool for forest monitoring and protection purposes. WSN deployment allows forest researchers to understand the impact of air pollutants ($CO_2$, ozone, etc.) and climate change on growth of trees. It also enables forest administration to formulate effective forest policies to ensure long-term health of forest ecosystem. We consider a use case in which a forest protection agency collects environment data using sensor infrastructure provided by a third party VWSN provider. The sensors are of various sensing capabilities and include temperature, humidity, rain-gauge, $CO_2$ detector, and wind speed sensors. Sometime later, the agency might be interested in sensor data of increased granularity with both space and time. Thus, it will use the infrastructure provided by another third party VWSN provider. Let us consider wildfire management agency that needs to be notified when fire occurs in the forest so that it can use countermeasures to suppress fire. The sensors of interest typically include temperature, humidity, and wind speed. WSN virtualization enables concurrent execution of forest monitoring application as well as wildfire management application on same sensors. In order to collect measurements from sensors, these applications need different gateway modules for handling different communication networks.

### B. Requirements

Our first requirement on the gateway is support for *standard and proprietary interfaces*. The proposed architecture must be *extensible* to support future scenarios and new application domains. Besides, the architecture must be *elastic* for efficient utilization of underlying physical resources. The architecture must be *scalable* to provide as many gateway modules as needed for a large WSN deployment. Scalability also promotes accelerated growth of number of applications without worrying about availability of the required gateway modules.

The architecture shall minimally provide key gateway modules: *protocol conversion function* and *information model processing*. Protocol conversion is needed for communication between different VWSN domains as well as communication between a VWSN domain and an application domain. In addition, involvement of multiple vendors and the eventual heterogeneity in WSN domain requires information model processing to parse the raw sensor measurements and encode in a specific format (e.g. XML, JSON, etc.) before sending it to the application.

The architecture must ensure that execution of gateway modules achieves similar *performance* compared to when they are executed in a traditional WSN gateway. In particular, the performance metrics that require significant attention include latency, throughput, and overhead. The NFV architecture must be *flexible* enough to support integration of sensors of different brands. Last but not least, the NFV architecture must have the *ability to support different business models*.

### C. The State-of-the-Art and its Shortcomings

Several WSN gateway architectures [6]-[10] have been proposed in the literature. Our motivating scenario closely resembles IoT scenarios which involve a broad range of IoT devices, several communication technologies at the IoT device domain (e.g., 6LowPAN, ZigBee, Bluetooth and RFID) as well as at the network domain (e.g., 2G/3G, LTE, LAN, and DSL). The state-of-the-Art for WSN/IoT gateway architectures has the common goal of bridging different sensor domains with public communication networks and internet. A growing trend towards NFV based middlebox design has been witnessed in the literature. Since, a WSN gateway falls under the taxonomy of middlebox, a brief overview of NFV architectures in the context of middleboxes is extremely important. We, therefore, classify the state-of-the-Art into the following two categories: Traditional Architectures (WSN/IoT gateway) and NFV architectures (middleboxes).

*1) Traditional Architectures (WSN/IoT Gateway):* An architecture for in-home IoT gateway is proposed in [6]. It consists of three subsystems: sensor node, gateway, and application platform. The sensor node collects sensor data and receives commands from gateway. The gateway provides communication modules to exchange data with a remote server. Application platform provides user control interfaces, configure and manages gateway as well as sensor network. This work has several distinguishing characteristics, however the elasticity of the architecture for efficient utilizaition of resources and the scalability in terms of number of applications is not discussed. Also, it doesn't provide support for standard as well as proprietary interfaces. Protocol conversion is partly satisfied. Lu et al. [7] present an IoT gateway architecture for CorbaNet based digital broadcast system. It is extensible in nature, but is limited in terms of protocol conversion abilities. A configurable, multifunctional and cost-effective architecture [8] is proposed for smart IoT gateway. Since modules with different communication protocols can be plugged in, the architecture satisfies the extensibility requirement. It also offers several unified user interfaces that allow external users to specify their requirements, avoid developing different types of gateways for different applications, and reduce the operational cost. The architecture addresses protocol conversions by introducing a common frame structure for data communication. Despite several important features, the possibility to use gateway modules (e.g. communication module) by a large number of simultaneous applications is not discussed. Thus, scalability is yet to be investigated. Besides, elasticity of provisioning gateway modules is not discussed.

In [9], an IoT gateway centric architecture is proposed to provide various M2M services such as discovery of M2M devices and endpoints (i.e. sensors and actuators) by mobile clients, and association of metadata to sensor and actuator measurements using Sensor Markup Language (SenML) representation. Since, SenML is lightweight, it allows efficient and fast parsing of metadata of a large *number of sensors*. *Although* being scalable in terms of handling traffic using RESTful paradigm, the architecture cannot allow dynamic creation of more M2M services (e.g. API to create desired SenML metadata) with increase in number of IoT devices and endpoints. In [10], gateway architecture for home and building automation system is proposed. It has support for standard and proprietary technologies which also allow it to extend the gateway capabilities. However, scalability and elasticity features are not discussed. Protocol conversion requirement is partly satisfied by the gateway architecture.

*2) NFV architectures (Middleboxes):* ClickOS [15] is a Xen based software platform which realizes NFV by allowing hundreds of middleboxes to run on commodity hardware. The important components in the ClickOS architecture are Xen's network I/O subsystem, middlebox virtual machines based on Click, and tools for building and managing ClickOS VMs. Although the ClickOS prototype includes both simple (e.g., forwarding packets from input to output interface, swapping Ethernet source and destination fields) and fullfledged middleboxes (e.g., IPv4 router, Firewall, NAT, etc), virtualizing gateway modules is not investigated. The architecture is scalable, flexible, extensible, and satisfies the performance requirements stated above. Ge et. al [14] identify that it is not feasible to obtain required performance for some of the middleboxes (e.g. Deep Packet Inspection [DPI], Network Deduplication [Dedup], and NAT), when executed on the commodity hardware. To address this issue, OpenANFV architecture is proposed in [14], wherein middlebox processing is accelerated by using virtualized FPGA card. Since VNFs can be created, migrated, and terminated on demand, OpenANFV satisfies elasticity. T-NOVA (Network Functions as a Service over Virtualized infrastructure) [12] is an integrated architecture which in addition to enabling network operators/service providers to manage their NFVs, also provides virtualized network functions ranging from flow handling control mechanisms to in-network packet payload processing as value-added services to operators' customers. T-NOVA allows third party developers to publish VNFs as independent entities which can be selected by customers and plugged into the customers' connectivity services. This feature enables customers to adapt network functions according to their needs. In [11], NFV is adopted for virtualizing Internet Protocol telephony function called Session Border Controller (SBC) function. An SBC operates at the edge of two separate networks, both on the control plane and on the media plane. On the control plane, it performs load balancing and call-control; on the media plane, the SBC provides media adaptation capabilities, i.e. it can adjust in real time the coding format of the speech signals transmitted by the users. NFV for virtualizing routing function in openflow enabled network is explored in [13]. An openflow controller is used to detect the packet received by a switch, analyze the information contained in the packet (i.e. type of protocol, source addresses, target address), and then send it to virtualized routing function, if needed.

From the above discussion, we conclude that existing WSN gateway architectures fall short of many of our requirements except limited support for extensibility, proprietary interfaces, and gateway modules. As evident for NFV based solutions, the current NFV architectures for middleboxes exhibit extensibility, scalability, and elasticity properties. However, they primarily focus on network elements such as firewall, proxies, NATs, etc., and middlebox processing from the perspective of a WSN gateway is not investigated so far.

## III. PROPOSED NFV ARCHITECTURE FOR VIRTUALIZED WSN GATEWAY

In this section, we present our NFV architecture for virtualizing WSN gateway. The architectural principles are discussed first, followed by the architectural components, interfaces, VNFs, VNF migration, and illustrative scenario.

### A. Architectural Principles

Our first architectural principle is that virtualized WSN gateway is decomposed into two key network functions: protocol conversion and information model processing. This decomposition leads to effective utilization of virtualized WSN gateway. Our second principle is that each network function has multiple VNF instances. This principle emerges from the fact that our case study assumes two different VWSN, each of which uses a specific brand of sensors. Our third principle is centralized store of VNFs, the images of which are stored in VWSN gateway provider. When the VWSN gateway provider receives request from a VWSN provider for a specific VNF, it instantiates and migrates the requested VNF images to VWSN domain. This principle is in accordance with the ETSI [2] that VNFs must be deployed throughout the networks, where they are most effective and highly customized to a specific application or user. VNF instantiation and migration as described above has several advantages. First, the time to boot or instantiate VM in VWSN domain is saved. Second, we eliminate the need to transfer large bulk of data from each VWSN domain to VWSN gateway provider domain. Also, migration of VM images that already exist in VWSN domain incurs lower delay compared to migration of a new VM image. The fourth and last principle is that VNFs responsible to realize a network service (e.g., sending sensor measurement to the forest monitoring application) are executed using a static chain. This design choice is motivated by the scenario covered in our case study, wherein the information model processor module

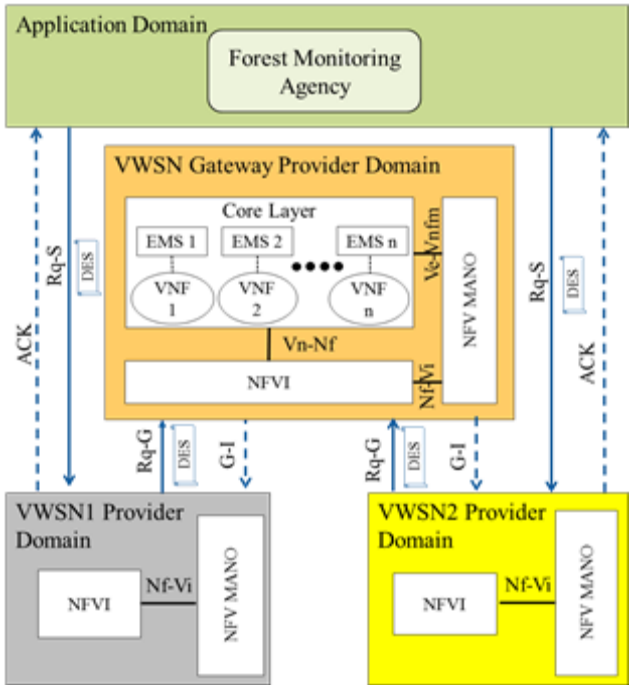

Fig. 1. Overall Architecture

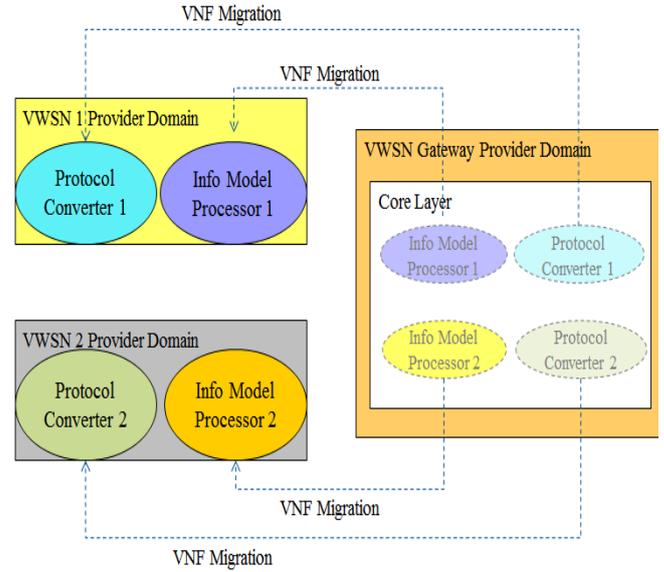

Fig. 2. VNF Migration

always needs to be executed before the protocol converter module.

### B. Overall Architecture

*1) Components:* Fig. 1 shows the proposed architecture. It is comprised of four domains: application domain, VWSN gateway provider domain and two distinct VWSN provider domains: VWSN1 and VWSN2. VWSN gateway provider domain has the following NFV components: core layer, NFV Infrastructure (NFVI), and NFV Management and Orchestration (MANO). Core layer is responsible for deployment of VNFs. Each VNF is managed by an EMS. NFVI provides the hardware and software resources including computation, storage, and networking necessary to deploy, manage, and execute VNFs. MANO comprises of orchestrator, VNF manager, and virtualized infrastructure manager. The orchestrator provides orchestration and management functions to realize execution of VNFs on NFVI. The VNF manager is responsible for managing life cycle of a VNF such as instantiation, update, scaling, and termination. The virtualized infrastructure manager provides functionalities to control interaction of VNF with computing, network, and storage resources as well as their virtualization.

Each VWSN provider provides an infrastructure for virtualization. Thus it comprises an NFVI for execution of VNFs and MANO to orchestrate and manage VNF life cycles.

*2) Interfaces:* For interaction among NFV components of our architecture, we adopt the interfaces defined by ETSI in [2]. They include Vn-Nf, Nf-Vi, and Ve-Vnfm. The interface Vn-Nf represents the execution environment provided by NFVI to VNF. It is not bound to a specific control protocol. The interface Nf-Vi is used for assigning virtualized resources (e.g. allocate VMs on hypervisors) in response to resource allocation requests. This interface is also used by NFVI to communicate state information of virtualized as well as hardware resources with the virtualized infrastructure manager which is a management entity in MANO. Besides, Nf-Vi is used for configuring hardware resources. The interface Ve-Vnfm is used for carrying out all operations during a VNF life cycle starting from its instantiation, scaling, update, and termination. It is also used for exchanging VNF configuration information. VNF state information necessary for network service life cycle management are also exchanged via Ve-Vnfm. Besides NFV interfaces, we introduce control interfaces for our architecture. Essentially, these interfaces are designed to enable communication between different domains. The control interfaces are as follows: Rq-S, Rq-G, G-I, and ACK. The request-for-sensor (Rq-S) interface and acknowledgement (ACK) interface are used for exchanging service related information between application domain and VWSN provider domain. The request-for-gateway (Rq-G) interface and gateway-instance (G-I) interface are used for exchanging VNF related information (e.g. VNF description) between VWSN provider domain and VWSN gateway provider domain. All control interfaces are based on RESTful Web services. Since, REST supports a wide range of resource description

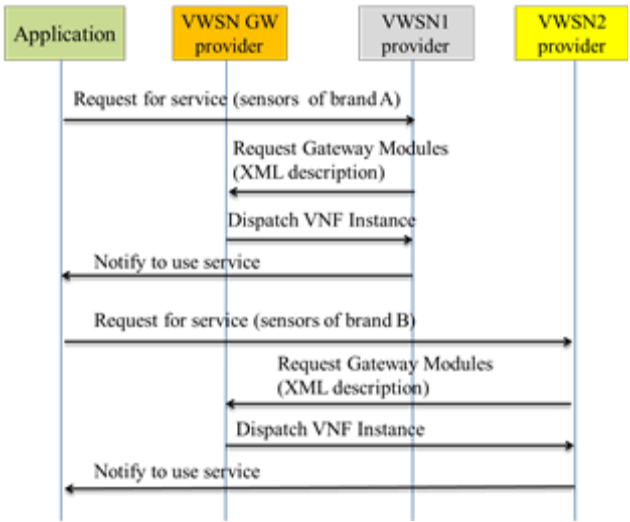

Fig. 3. Service Initiation

mechanisms (XML, JSON), VNFs can be described effectively using a suitable format.

*3) VNFs:* VNFs are deployed for various gateway modules, e.g. protocol converter and information model processor. Protocol converter decodes a packet received in one protocol and re-encode it in another protocol. Information model processor transforms sensor measurement from one format to another. In order to realize a specific network service, these VNFs need to be chained. In our architecture, VNFs are executed using a pre-defined or static chain. In this case study, information model processor is executed first followed by execution of protocol converter. This case study motivates us to design four VNFs: protocol converter 1 and information model processor 1 for VWSN1 provider domain, protocol converter 2 and information model processor 2 for VWSN2 provider domain. A static chain is decided for each VWSN provider domain.

*4) VNF Migration:* All VNFs are instantiated in the core layer of NFV in the VWSN gateway provider domain. They are migrated on-demand to a VWSN provider domain based on necessity. The instantiation and migration is orchestrated by MANO of VWSN gateway provider domain. Once a VNF is migrated to VWSN provider domain, its execution in the virtual environment is orchestrated and managed by the MANO of VWSN provider domain. If the number of services grows in VWSN provider domain, it requests the core layer to instantiate and migrate more VNF instances. Fig. 2 shows migration of VNF instances of protocol converter and information model processor.

### C. Illustrative Scenario

In this section, we illustrate a scenario which involves collection of sensor measurements by an end-user application (e.g. forest monitoring application). The service (i.e. collection of sensor measurements) is provided by VWSN1 provider as well as VWSN2 provider that own sensors of brand, let say A and B respectively. Before using the service, the forest monitoring application needs to ensure that both providers *have* accommodat*ed* the resources necessary to execute the task. Thus, the former sends a service request to VWSN1 provider using Rq-S interface. The request includes description of service such as sensing parameters, type of collection pattern (periodic or just once), etc. On receiving request, VWSN1 provider generates a request including description of the VNFs (i.e. virtualized gateway modules) to VWSN gateway provider over Rq-G interface. VWSN gateway provider spawns the required VNFs and dispatch*es* the same to the NFV infrastructure in the VWSN1 provider domain. *Once* required VNFs are deployed, the VWSN1 provider communicates with the end-user application over ACK interface notifying the latter to start using the service. Using *a* similar procedure, the end-user application is notified by VWSN2 provider. The sequence of operations from service request initiated by application till notification by VWSN providers, is depicted in Fig. 3. .

### IV. IMPLEMENTATION

### A. Implementation Setup

As a prototype, we implemented the forest monitoring scenario. We consider a forest where wireless sensor networks have already been deployed to monitor wildfires. The forest monitoring agency is interested in collecting environmental data to monitor the forests. Two different brands of sensors were used, each belonging to different WSN cloud infrastructure. The sensors can measure the temperature and thereby detect the fire. In order to communicate with different types of sensors, the agency needs a gateway for handling different types of communications. This gateway is provided by third party provider.

We used six Java SunSpots sensors and two advanticsys sensors. Each sensor executes the forest monitoring task. The application task is coded in Java 2 Platform Micro Edition (J2ME). J2ME is a robust and flexible Java platform that enables the development of applications for mobile and embedded devices. A RESTful web service is used by the forest monitoring agency node to receive sensor measurements. The sensors send raw measurements using HTTP. The raw data received from Java SunSpot and advanticsys is mapped to JSON format according to SenML specifications. The combination of SensorML and O&M is another option, but we selected SenML since it is less complex. The forest monitoring application is created using regular Java web application. The forest monitoring application runs on a laptop with an Intel Core i5 CPU clocked at 2.67 GHz, and a 4GB RAM with 32bit Windows 7 Enterprise. This laptop used JVM version 1.7.0_21.

### B. Implementation Architecture

The implementation architecture is shown in Fig. 4. It is comprised of two domains: VWSN gateway provider domain and VWSN domain. We configure Openstack at both domains. The Openstack components in VWSN gateway provider domain are as follows: orchestrator ("Heat"), identity service ("Keystone"), compute ("Nova"), image ("Glance"), object storage ("Swift"), block storage ("Cinder"), and networking ("Neutron"). The VWSN domain contains compute and image components. Compute component of OpenStack stores and retrieves virtual disks and associated metadata in image component. Network component provides virtual networking for compute. Block storage provides storage volumes for compute. Image can store the actual virtual disk files in the object store. Identity service provides authorization for all other components. All interactions between OpenStack components in VWSN gateway provider domain and that in VWSN domain are achieved through OpenStack Rest APIs. We build VM images for the following VNFs: protocol converter 1, information model processor 1, protocol converter 2, and information model processor 2. The implementation architecture also supports migration of VM instances from VWSN gateway provider domain to VWSN domain. To be specific, VM instances of protocol converter 1 and information model processor 1 are migrated to process data generated by sun spot sensors; whereas VM instances of protocol converter 2 and information model processor 2 are migrated to process data generated by advanticsys sensors.

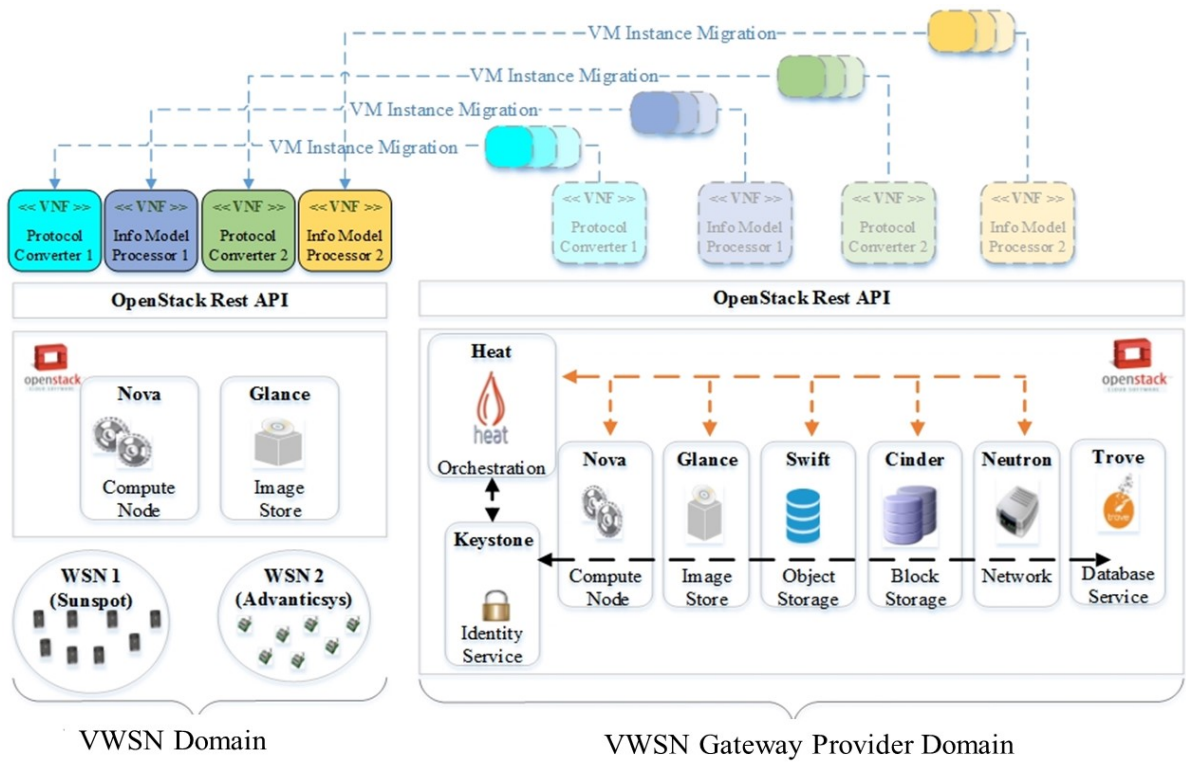

Fig. 4. Implementation Architecture

## V. CONCLUSION AND FUTURE WORKS

The case study presented in this paper shows the usefulness of NFV deployment in VWSN environments. We show that NFV makes it possible the dynamic, elastic, and scalable deployment of gateway modules in heterogeneous VWSN environment. Besides, involvement of several business actors opens the door for unique business models that leverage NFV to reducing operational as well as maintenance cost.

In future works, we plan to address dynamicity in service chaining in our NFV architecture. Also, we extend the architecture to other potential domains for virtualization such as robotic networks. We also plan to address service level agreement.

## ACKNOWLEDGMENT

This work is partially supported by by CISCO systems through grant CG-576719.